# Large-Flip Importance Sampling

**Firas Hamze**
Computer Science, UBC
Vancouver, Canada

**Nando de Freitas**
Computer Science, UBC
Vancouver, Canada

## Abstract

We propose a new Monte Carlo algorithm for complex discrete distributions. The algorithm is motivated by the N-Fold Way, which is an ingenious event-driven MCMC sampler that avoids rejection moves at any specific state. The N-Fold Way can however get "trapped" in cycles. We surmount this problem by modifying the sampling process. This correction does introduce bias, but the bias is subsequently corrected with a carefully engineered importance sampler.

## 1 Introduction

Sampling from densely connected discrete probabilistic graphical models is a fundamental problem in physics, statistics and artificial intelligence (AI). In AI, the samples are typically used to infer the values of unknown random variables or to compute the partition function for the purposes of parameter and model selection. Obvious application domains include Boltzmann machines [19], densely connected conditional random fields [15] and densely connected graphs arising in semi-supervised learning [9]. These problems are notoriously hard to solve. A less obvious application domain arises when dealing with models with discrete and continuous variables, where the continuous variables can be integrated out analytically. This domain is very broad and of great applicability. We give a few examples of members of this class subsequently.

The first example is mixture models. For example, in a mixture of Gaussians, one can integrate out the mixture proportions, means and variances and end up with an un-normalized "densely connected" distribution, which depends only on the discrete mixture component indicator variables [16]. After sampling these indicators, one can compute the other variables analytically. This is often referred to as Rao-Blackwellization or Collapsed Gibbs sampling. Likewise, in latent Dirichlet allocation [2] one also ends up with a distribution in terms of indicators (involving ratios of Gamma functions) after carrying out analytical integration. A final example to illustrate this point is variable selection with kernel regression models [24]. Here, after integrating out the kernel coefficients and noise covariance, one is left with a densely connected distribution over the binary variables indicating the presence of features.

In all these densely connected models, sampling the discrete variables is well known to be a very demanding task. Variational methods also tend to exhibit estimation problems in these domains [4]. The target distributions in these models tend to have many modes separated by low probability barriers that cannot be crossed by standard moves. MCMC methods often get "stuck" in modes of the target distribution and fail to accurately sample from all relevant regions in the time allotted to the experiment. For example, as reported in [18], it could take $10^{10}$ minutes for a dynamic Metropolis algorithm to leave a metastable state in an Ising model.

To surmount this problem, approaches based on ingenious domain specific proposal distributions have been proposed. We are however interested in developing general methods. With this goal in mind, the most successful methods seem to be cluster techniques, such as the Swendsen Wang algorithm [23], population-based methods and multicanonical Monte Carlo [12, 17]. Theoretical studies have shown that the so-called *mixing time* can diverge exponentially in the Swendsen-Wang algorithm [10]. Naive parallel chain methods are often wasteful and computationally expensive [12]. Multicanonical methods require the estimation of the density of states and this is no easy task; see [12] for an excellent review.

Distinct from plain parallel MCMC are the *Sequential Monte Carlo,* methods [13, 6] that use importance sampling in a joint, artificially-imposed temporal state-space to compensate for the effect of incomplete convergence of the proposal processes. Unfortunately, such methods are also limited by the proposal processes used, and can only do as well as importance sampling with the (temporal) marginal distribution of the set at the last step of the sequence. In general this marginal is not known, so weighting is done in the joint temporal space, incurring additional variance. While for systems with a relatively simple state-space, say with relatively few but widely-separated modes, these methods can yield great improvements, when the target space becomes very rough, we expect the situation to degenerate. In the former case, there are few "obstructions" in the way of the particles' movement towards the high density regions, though once they reach them leaving may take a long time. In contrast in the latter case, which occurs in Bayesian modelling, statistical physics and biopolymer simulation, the number of particles is tiny compared to the number of modes that can trap them, and so it again takes an exponentially long time to reach the "important" regions.

In this paper, we proposed a novel approach to this problem. It is motivated by the N-Fold Way sampler proposed in [3]. This sampler has an ingenious way of avoiding the situation of being trapped in any specific state. It can however get easily trapped in cycles. In our approach we surmount this problem partially by forbidding the chain from jumping to recently visited states. This process is no longer Markov and introduces bias. We correct for the bias by adopting a mixture of Gibbs proposal distributions, initialized at our biased process, within an importance sampler with the right target distribution. This approach is distinct from methods traditionally known as "importance sampling" for Bayesian Networks in the UAI community, which, for example, weight samples from the prior distribution with the likelihood [8], or adaptively construct an approximation to an optimal proposal [5]. In particular, the model need *not* be specified as a product of known conditional distributions.

## 2 Preliminaries and Notation

We are interested in sampling from the following target distribution:
$$\pi(x) = \frac{e^{-\beta E(x)}}{Z(\beta)}. \quad (1)$$

In general we can only evaluate $\pi$ up to a constant and refer to this unnormalized function as $\tilde{\pi}$. Here, $E(\cdot)$ denotes the energy, $\beta$ the inverse temperature and $Z$ the partition function.

We assume the following in the problems we consider. First, that $x$ is a $M$ dimensional vector, each component of which can assume $q$ values in the set $\mathcal{D} \triangleq \{D_1 \ldots D_q\}$, so that the state space $\Omega$ of $\pi$ is of size $M^q$. In the sequel, we often refer to different values of $x$ in a sequence, in which case $x_{t,i}$ means the $i^{th}$ component of $x_t$. Second, we can readily calculate and sample from the conditional distributions of $\pi$. If the set of all variables is $\mathcal{M} \triangleq \{1, 2, \ldots, M\}$, then the conditionals $\pi_i(x_i|x_{\{\mathcal{M}\setminus i\}})$ are available in that sense. Here, $x_{\{\mathcal{M}\setminus i\}} \triangleq (x_1, \ldots, x_{i-1}, x_{i+1}, \ldots, x_M)$. In this paper we assume that readers are familiar with constructing and sampling from the local conditionals of models with distributions of the form (1) such as Markov Random Fields (MRFs.)

Our tasks are *inference,* the computation of quantities expressible as
$$E_\pi[h(X)] \quad (2)$$
and *normalization,* the estimation of the unkown constant
$$Z(\beta) \triangleq \sum_x e^{-\beta E(x)} \quad (3)$$

Marginalization is a special case of Equation (2). Needless to say, obtaining exact solutions to these quantities is out of the question in general for realistically-sized models.

## 3 Event-Driven Monte Carlo

We motivate our algorithm by first considering a beautiful restructuring of traditional single-variable MCMC algorithms that can result in dramatic simulation efficiency gains. Algorithms using this methodolgy comprise a family known as *Event-Driven Monte Carlo,* of which the first, due to [3], is known as the *N-Fold Way* (NFW). Although our approach diverges from the spirit of these methods, particularly by forsaking any interest in having a proper MCMC process, it is worth looking at them in some detail since the exhibition gives some insight into the problems that arise while running MCMC and how our approach attempts to circumvent them.

Traditional single-site MCMC, say using a Gibbs or Metropolis process, proceeds by choosing a site at random and sampling a new value according to a local transition probability. In this work we focus on using the Gibbs sampler without loss of generality; adaptation to the (plain) Metropolis method is straightforward. The states which are the same as a given state $x$ but for one component are called the *single-site neighbours* of $x$. Formally, the random-site Gibbs sampler transition kernel $K_G(X_t = x_t | X_{t-1} = x_{t-1})$ is defined as:

$$\frac{1}{M} \sum_{i=1}^{M} \pi_i(x_{t,i}|X_{t-1,\{\mathcal{M}\setminus i\}} = x_{t-1,\{\mathcal{M}\setminus i\}})$$
$$\times \delta_{(x_{t-1,\{\mathcal{M}\setminus i\}})}(X_{t,\{\mathcal{M}\setminus i\}}). \quad (4)$$

The kernel is an equal-weight mixture of update densities. The delta functions in each mixture component are to specify that $X_{t,\{\mathcal{M}\setminus i\}} = X_{t-1,\{\mathcal{M}\setminus i\}}$; $X_{t,i}$ is sampled from the local conditional.

If the current state is such that all potential neighbors are energy-gaining moves, then at low temperatures, $X_t$ will very likely be $x_{t-1}$; the system is forced to "wait" in the same state until the low-probability event that one of the uphill neighbors is transitioned to.

This perspective on the single-site MCMC motivates the NFW. If one can calculate the probability that a variable will change, then instead of repeatedly taking MCMC steps, we can sample a waiting time $\tau$ from this probability, and then sample from the *conditional* distribution of the neighboring states *given that a change occurred*. This gives a statistically equivalent process to the conventional MCMC sampler; the state $x_{t-1}$ can be "replicated" for $\tau$ steps and a new state $x_{t+\tau}$ is then sampled from this conditional.

We now present the NFW more specifically. First, for notational compactness, define:

$$\zeta_i(x'; x_{t-1}) \triangleq \frac{1}{M} \pi_i(x' | X_{t-1, \{\mathcal{M} \setminus i\}} = x_{t-1, \{\mathcal{M} \setminus i\}}). \tag{5}$$

$\zeta_i(x'; x_{t-1})$ is simply the probability that variable $i$ is selected uniformly at random and receives sampled value $x'$ (which could, of course be the same as $x_{t-1,i}$); all other variables stay as they were at $t-1$. The problem of excessive waiting time referred to previously occurs when for each $i \in \mathcal{M}$, $\zeta_i(x'; x_{t-1})$ is overwhelmingly peaked on $x_{t-1,i}$, while the remaining $q-1$ values that $X_{t,i}$ can assume are all unlikely. Let the set of values that variable $i$ can assume if it changes value from $x_{t-1,i}$ be $\mathcal{F}_i(x_{t-1}) \triangleq \{\mathcal{D} \setminus x_{t-1,i}\}$. We will sometimes refer to the operation of changing a variable's value as *flipping* it, though the term is best-suited to the binary ($q=2$) case.

Now let

$$\alpha_i(x_{t-1}) \triangleq \sum_{x'_{t,i} \neq x_{t-1,i}} \zeta_i(x'_{t,i}; x_{t-1}). \tag{6}$$

That is, $\alpha_i(x_{t-1})$ is the probability that variable $i$ is chosen by the Gibbs sampler *and is changed* from its value of $x_{t-1,i}$; thus for each $i \in \mathcal{M}$ there are $q-1$ terms in the summation (6). Once again, in a local minimum and at high $\beta$, these probabilities are low for all $i$. Now define the *marginal probability that a change occurs* during a Gibbs step to be:

$$\begin{aligned} p_{flip}(x_{t-1}) &\triangleq Pr(X_t \neq x_{t-1}) \\ &= \sum_{x_t \neq x_{t-1}} K_G(x_t | X_{t-1} = x_{t-1}). \end{aligned} \tag{7}$$

It is clear that $p_{flip}(x_{t-1}) = \sum_{i=1}^M \alpha_i(x_{t-1})$. The event that a change in state occurs can then be seen as the first occurence of a "success" in a series of Bernouilli trials with success probability $p_{flip}(x_{t-1})$; the time until such an event, is of course, distributed according to a *geometric distribution*, from which samples can be drawn very efficiently [7].

Finally, let us create a discrete distribution out of the set of $(q-1)M$ values that the $\zeta_i(x'; x_{t-1})$ can assume for $i \in \mathcal{M}, x' \neq x_{t-1,i}$. Let

$$\hat{\nu}(i, x; x_{t-1}) \triangleq \begin{cases} \zeta_i(x; x_{t-1}) & \text{if } x \in \mathcal{F}_i(x_{t-1}) \\ 0 & \text{otherwise} \end{cases} \tag{8}$$

and

$$\nu(i, x; x_{t-1}) \triangleq \frac{\hat{\nu}(i, x; x_{t-1})}{\sum_{j \in \mathcal{M}, x' \in \mathcal{F}_j(x_{t-1})} \hat{\nu}(j, x'; x_{t-1})} \tag{9}$$

The reader should convince herself that $\nu(i, x; x_{t-1})$ is the posterior probability of the joint event "variable $i$ assumes value $x$" *given that a change in state occurred*, *i.e.* when $x_t \neq x_{t-1}$. The posterior will favor sampling the "best" of the neighbors of state $x_{t-1}$, i.e. those resulting in the smallest energy gain. At high $\beta$, these will be chosen the vast majority of the time. Sampling from this distribution is $O(M(q-1))$ in the worst case, though many optimizations are possible in certain cases, such as graphical models of sparse connectivity or when the set of possible moves can be placed into one of a small set of classes, as was done in the original paper of [3]. In this work we describe experiments on systems where such optimizations are not possible as far as we could tell, namely densely-connected graphical models with continuous-valued *parameters*, so the optimizations will not be discussed here.

Thus, one possible statement of the NFW is as follows. First, choose a number $T-1$ of flip moves to make, so that the process including the initial state has $T$ steps. Let us define a process $\hat{X}_0, \ldots, \hat{X}_T$, each having the same domain as the original problem. The indices here correspond to *flipping* moves and *not* the "real" time of the simulation.

$\hat{X}_0$ is the initial (randomly chosen) state, and $\hat{X}_n$ is sampled from $\nu(i, x; x_{n-1})$. Furthermore, define a sequence of *time* variables $\Theta_0, \ldots, \Theta_T$. The algorithm

---

1. Set $X_0 = \hat{X}_0$ and $\Theta_0 = 0$.
2. For $i = 1 \ldots T$
   (a) Sample $\tau \sim Geometric(p_{flip}(x_{i-1}))$
   (b) Sample $\hat{X}_i \sim \nu(i, x; x_{i-1})$
   (c) Set $\Theta_i = \Theta_{i-1} + \tau$

---

Figure 1: N-Fold Way sampler.

is constructed as shown in Figure 1. The states in Monte Carlo time can then be read off from this data: $X_{\Theta_i \ldots \Theta_{i+1}-1} = \hat{X}_i$. The effective length of the simulation, *i.e.* that of an equivalent direct Gibbs sampler, is $\Theta_T$. The overall complexity of the method without any optimizations is $O((q-1)M)$ per step, and is thus $O((q-1)MT)$ for all $T$ steps.

It should also be mentioned that one can apply this strategy to the *simulated annealing* [14] optimization heuristic by simply replacing the local conditionals $\pi_i(x | X_{n-1, \{\mathcal{M} \setminus i\}} = x_{n-1, \{\mathcal{M} \setminus i\}})$ in the preceeding calculations with their annealed versions $\pi_i^{\gamma_n}(x | X_{n-1, \{\mathcal{M} \setminus i\}} = x_{n-1, \{\mathcal{M} \setminus i\}})$ where $\gamma_n$ is the temeperature parameter chosen to "cool" according to some schedule. Note that in this event-driven formulation, any annealing schedule is effectively such that in a conventional implementation, the temperature is held constant at $\gamma_n$ between the times of the $(n-1)^{\text{th}}$ and $n^{\text{th}}$ flips. Also in most optimization contexts, one is not interested in the times $\Theta_i$ of the flips; the method reduces to sampling from annealed versions of (9), which effectively performs a stochastic best-improving search that becomes less "soft" as the annealing continues.

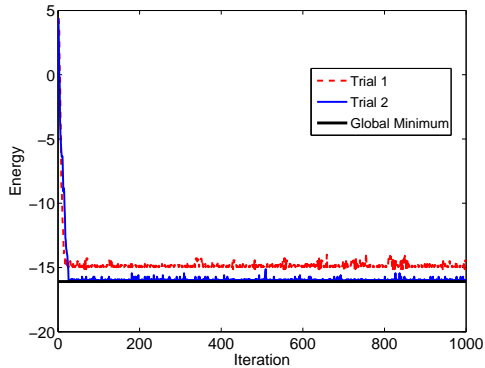

Figure 2: A binary-valued system evolving under two different runs of the N-Fold Way at $\beta = 5$. Once the processes find a local basin, they can take excessively long to escape.

The need for computing the flip (or equivalently rejection) probabilities in the NFW creates difficulties when considering continuous variables. Often these are not integrable. However for simple continuous distributions, some progress has been made [18].

While the NFW is an example of how a clever perspective can yield great practical improvements, if it solved the problem of simulating from complex distributions, then this paper need not have been written. A severe shortcoming might have suggested itself to the reader in the description of sampling from the conditional (9). If this conditional favors choosing the best of the (bad) neighbors in a local minimum, then unless that move results in favourable moves other than the original local minimum, the next sample from (9) will almost certainly re-select the original local minimum since it is now a favourable point from the new one. Breaking out of this cycle can take an extremely long time. An example of this behavior is shown in Figure 2; the simulation shows evolution of the energy of a 25 node fully-connected binary spin glass with Gaussian-generated parameters at the low temperature $\beta = 5$. (this model will be detailed in a coming section.) Note how two different runs of the algorithm get stuck in minima of different values, a very undesirable property of a Monte Carlo method.

A conceivable remedy to this problem is to include larger than single-flip neighborhoods into the algorithm. Indeed it may be possible that flipping a pair or a triplet can escape the minima. Unfortunately, while this is certainly conceptually correct, the computational costs make it infeasible. Even in the $q = 2$ case, there are $\binom{M}{k}$ sets of variables of size $k$ to flip, and these must all be included in the conditional (9) For small $k$ and large $M$ the strategy may not be worth the effort as quite large flips of state are often required; for large $k$ calculating and sampling from (9) is intractable.

In the next section, we finally come to our contribution to this problem in an attempt to overcome shortcomings of the NFW.

## 4   The Large-Flip Quasi-Gibbs Sampler

The low-temperature cycling behavior is a severe drawback of the NFW. As we mentioned, it is possible to approach low-energy states slowly via annealing into the desired distribution, i.e. using a sequence of distributions of the form $\pi^{\gamma_n}$ rather than having the sampler target the desired distribution $\pi$ from the beginning.

Our strategy is philosophically different and involves introducing a "memory" into the process. The resulting sequences of random variables no longer form Markov Chains due to the dependence on the history, and no longer (asymptotically) draw samples from $\pi$ in the same sense that MCMC methods do. However they do aggressively visit the important regions of $\pi$, even at very low temperatures, just not with the correct frequencies. This deficiency is subsequently corrected.

First, we drop interest in the "real" time of the process, i.e. the sequence $\Theta_i$, since this is only useful for inducing a correct MCMC process, which we will not have anyway. Our sequence of samples $X_n$ corresponds to the $n^{\text{th}}$ *flips* from the initial state $X_0$. However rather than choose flips according to the conditional distribution (9), the new mechanism is such that once a variable has been chosen for change, it is prevented from re-assuming its previous value for a certain amount of simulation time. (From now on unless otherwise clarified, "time" refers to the sequence of flips rather than MCMC time.) The reader may sense a connection between this idea and a much-used meta-heuristic for difficult combinatorial problems known as *Tabu Search* [20]. In fact we were initially motivated to solve the problems faced by the NFW, but this connection became made clear to us. Our subsequent survey of the tabu search literature has motivated one aspect of the algorithm, namely the random choice of set sizes of variables to update. However strictly speaking, our method has no "tenure" in the same sense, nor any "aspiration" criteria or any of the other features of the heuristic.

Define the $k^{\text{th}}$ partial *Large-Flip* (LF) move to be the

(ordered) *sequence*

$$\chi_m^k \triangleq ((i,x)_0, \ldots, (i,x)_{m-1}) \quad (10)$$

The $k^{\text{th}}$ LF move is *complete* when the sequence $\chi_m^k$ is of length $\Gamma_k$, the *size* of the $k^{\text{th}}$ LF move. The elements $(i,x)_j$ of the sequence $\chi_{\Gamma_k}^k$ specify that variable $i$ has changed value to state $x$ at the $j^{\text{th}}$ step of the move, and thus implicitly determine the global state trajectory. For precision, we also define the (unordered) *set* of moves in the $k^{\text{th}}$ partial LF move to be:

$$\mathcal{C}_m^k \triangleq \{(i,x)_0, \ldots, (i,x)_{m-1}\} \quad (11)$$

If we define $\Gamma_{-1} = 0$, the cumulative number of flips during the $m^{\text{th}}$ step of the $k^{\text{th}}$ LF move is $n(m,k) = m + \sum_{i=0}^{k-1} \Gamma_k$; whenever we mention the global flip index $n$ this dependence on $m, k$ is to be remembered. The *onset time* of the $k^{\text{th}}$ LF move is $T_k^C = \sum_{i=0}^{k-1} \Gamma_i$ with $T_0^C = 0$.

Our method of escaping the cycling traps corresponds to forcing the probability of any duplicate elements $(i,x)_j$ within each complete LF move to be zero.

This requires a modification of equation (8); the goal is achieved with

$$\hat{\nu}(i, x; x_{n-1}) \triangleq \begin{cases} \zeta_i(x; x_{n-1}) & \text{if } (i,x) \notin \mathcal{C}_m^k \\ 0 & \text{otherwise} \end{cases} \quad (12)$$

When normalized as in equation (9), this will still result in a biased selection towards the "good" neighbors of state $X_{n-1}$, but such moves are suppressed if they have been visited during the current LF move.

The sizes of the LF moves $\Gamma_k$ are randomly chosen integers from an interval $[\hat{\Gamma}_{min}, \hat{\Gamma}_{max}]$ prior to the onset of the $k^{th}$ move. These limits are tunable parameters of the system and are chosen a priori. Our experiments with large models have used $\hat{\Gamma}_{max} = \lfloor M/6 \rfloor$ and $\hat{\Gamma}_{min} = \lfloor M/8 \rfloor$. The NFW flip selection equation (9) is the special case of $\Gamma_{min} = \Gamma_{max} = 1$. It is this random LF move size that was motivated by *robust tabu search*, a heuristic that has shown promise on the quadratic assignment problem [22] and MaxSAT [21].

While we do not care about the waiting time between samples, we will need to store the entire set of states visited by the process, which can be implicitly done in a memory-efficient way by remembering only the initial state $x_0$ and the entire set of LF moves, which only consist of (variable, value) pairs.

The pseudocode to this process, which we call the *Large-Flip Quasi-Gibbs Sampler* (LFQGS) is shown in Figure 3. The pseudocode contains many steps that are purely illustrative and would never be implemented that way; implementation decisions must always be made specifically to the problem at hand.

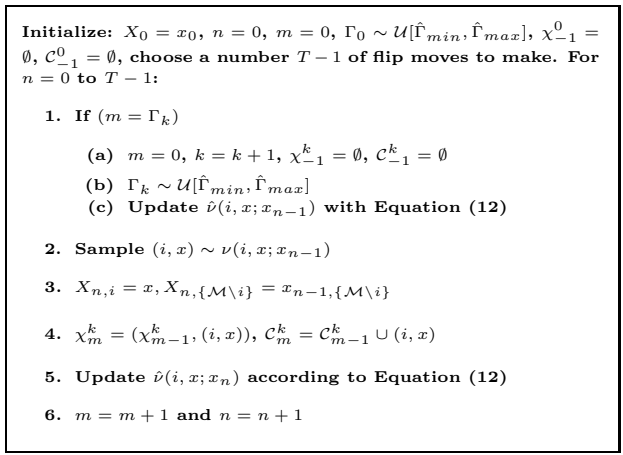

**Initialize:** $X_0 = x_0$, $n = 0$, $m = 0$, $\Gamma_0 \sim \mathcal{U}[\hat{\Gamma}_{min}, \hat{\Gamma}_{max}]$, $\chi_{-1}^0 = \emptyset$, $\mathcal{C}_{-1}^0 = \emptyset$, choose a number $T-1$ of flip moves to make. For $n = 0$ to $T-1$:

1. **If** $(m = \Gamma_k)$
   (a) $m = 0$, $k = k+1$, $\chi_{-1}^k = \emptyset$, $\mathcal{C}_{-1}^k = \emptyset$
   (b) $\Gamma_k \sim \mathcal{U}[\hat{\Gamma}_{min}, \hat{\Gamma}_{max}]$
   (c) **Update** $\hat{\nu}(i, x; x_{n-1})$ **with Equation (12)**

2. **Sample** $(i, x) \sim \nu(i, x; x_{n-1})$

3. $X_{n,i} = x$, $X_{n, \{\mathcal{M} \setminus i\}} = x_{n-1, \{\mathcal{M} \setminus i\}}$

4. $\chi_m^k = (\chi_{m-1}^k, (i,x))$, $\mathcal{C}_m^k = \mathcal{C}_{m-1}^k \cup (i,x)$

5. **Update** $\hat{\nu}(i, x; x_n)$ **according to Equation (12)**

6. $m = m+1$ and $n = n+1$

Figure 3: Large-Flip Quasi-Gibbs sampler.

A moment's reflection will reveal that this algorithm does *not* produce correct samples from $\pi$. The process is not Markov (due to the dependence via the LF memory) nor is it stationary. Indeed, we observe that forcing variables to not take certain values implicitly introduces a change to the model being simulated; for example by introducing "infinitely strong" (deterministic) local evidence in a graphical model about a variable's values. Thus at any step, the flip selection corresponds to a legitimate MCMC algorithm *for that implicitly defined model* but at the next step, the flipped variable's values will be constrained and so the model has changed yet again! The reader may question our sanity in making such a bizarre proposal to a respectable conference such as UAI; hopefully the next section will restore their faith.

## 5 Correcting the Bias with Importance Sampling

Our answer is that we can indeed make the samples drawn from LFQGS a component of a statistically correct methodology for inference and normalization, that is approximating integrals under the stationary distribution, but we cannot have the system exhibit "real" dynamics as MCMC does. We call this methodology *Large-Flip Importance Sampling.*(LFIS) In a single LFQGS simulation of $T$ samples, $T-1$ flips, there will be a set $\mathcal{S}$ of *distinct* states visited, with $|\mathcal{S}| \leq T$. Many of these states have very low probability under $\pi$, but are necessarily traversed using the LF moves in order to escape the local optima. The approach we take is to run a set of $N$ such LF processes of $T-1$ flips each resulting in random variables $\{X_n^l\}$ for $n \in \{0, \ldots, T-1\}, l \in \{0, \ldots, N-1\}$.

One possibility is to assign each state $x_i$ in the set of *distinct* visited states $\mathcal{S}^l$ for $l \in \{0, \ldots (N-1)\}$ a

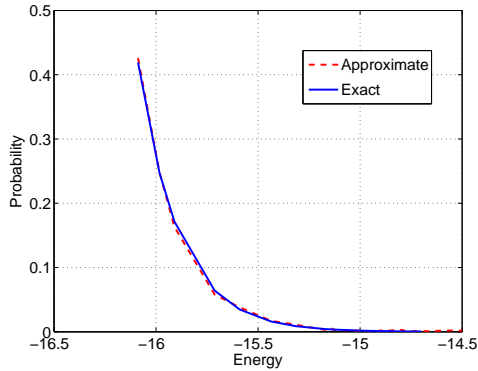

Figure 4: The exact and approximate probability distribution of the energy on a 25 node complete graphical model at $\beta = 5$. The approximation was calculated using the LFQGS followed by a selection step, as described in the text. The two plots are very close to each other.

weight:

$$q_i = \frac{e^{-\beta E(x_i)}}{\sum_{x' \in S^l} e^{-\beta E(x')}} \quad (13)$$

Selecting a state according to (13) corresponds to drawing an exact sample from the conditional distribution $\pi(x|x \in S^l)$, *not* from the full target $\pi(x)$. It bears a resemblance to the resampling step performed in Sequential Monte Carlo, though it is not the same as we are not selecting samples from the ratio of $\pi$ with a known proposal distribution. It should be emphasized right now that this feature plainly distinguishes the method from ones based on annealing; while in annealing the idea is to try to (slowly) move to a state of high probability, so that the final value is the one that is of primary interest, in LFQGS the selection method can concievably choose a sample visited at *any* point in the sequence $(X_0^l, \ldots, X_{T-1}^l)$, and it is indeed often the case that early samples are selected.

The selection process is only asymptotically correct in that as the number of samples $T \to \infty$, due to the randomness in the LF move sizes and the noise in the system, an ever-increasing portion of the overall state space is visited, so $\pi(x|x \in S^l)$ essentially becomes $\pi(x)$. This fact is of little practical consequence; what is desired is to collect a representative set of important states as early as possible.

For small systems, or large systems at low temperature, selecting according to (13) seems to accomplish this quite well. To illustrate, Figure 4 shows the exact probability distribution of the energies of a complete 25-node graphical model with binary values and many competing interactions at a temperature of $\beta = 5$ (see Section 6 for the model description) and the approximation obtained using the LFQGS followed by selection using $N = 1000$, $T = 1000$. The approximation was calculated by simply histogramming the energies of the $N$ selected states and took about 10 seconds of computer time on a 3GHz Xeon CPU. In contrast the brute-force computation of the exact distribution by summing over $2^M$ states took about 8 minutes on the same machine. The two curves are indeed very difficult to tell apart in Figure 4. We remind the reader again that due to the non-stationarity of the LFQGS process, it is *not* correct to conclude anything about $\pi$ by histogramming the samples $\{X_n^l\}_{n=0}^{T-1}$ in sequence $l$ as is done in MCMC.

For large systems at moderate temperatures, the simple weighting scheme above oversamples the low-energy states. For such cases, a more elaborate weighting is required; see [11] for a detailed discussion. Alternatively, one could simply apply a some extra iterations of the plain NFW to the selected samples.

Note that approximating the distribution of energies at $\beta = 5$ is demonstrating the method in a difficult case: the task is to generate samples with the correct nontrivial frequencies, not merely to find the global optimum, which is the case at very high $\beta$, or to in effect sample uniformly (low $\beta$).

The final piece of the algorithm allows us to estimate the partition function $Z(\beta)$ and to demonstrate its asymptotic correctness by appealing to standard importance sampling arguments. The set of operations described previously, that is the LFQGS followed by selection, effectively draws samples from some very complicated and unknown marginal distribution $f_T(x)$, which we hope, should be a reasonable approximation to $\pi(x)$. It is not asymptotically correct in the sense that as we draw $N \to \infty$ samples $Y^{(i)}$ from it for a fixed $T$, estimators of the form $\frac{1}{N}\sum_{i=1}^N h(Y^{(i)})$ do not converge to $E_\pi[h(X)]$ because $f_T(x) \neq \pi(x)$. Our solution to this is as follows.

Suppose we simulated $N$ LFQGS sequences for $T-1$ flips each. Define the set of $N$ selected variables under the rule (13) to be $Y^{(i)}$. For each $Y^{(i)}$, perform one sweep of *conventional* Gibbs Sampling through all the variables *in a pre-determined, fixed order*. The Gibbs Sampler will technically "move" the set of samples towards the stationary distribution $\pi$, whose unnormalized expression, we remind the reader, is $\tilde{\pi}$. This moving effect is slight and not the object of the exercise; using a Gibbs kernel just ensures that it will not make the population of $Y^{(i)}$ any worse. The application of the Gibbs kernel to the population $Y^{(i)}$ sampled from $f_T(.)$ results in samples $\tilde{Y}^{(i)}$ from a distribution we call $\mu(.)$. Since we can analytically compute $K_G(y|Y^i)$ due to the predetermined variable sweep (it is simply the product of the local conditional probabilities of the variables in the sweep order), it is straightforward to approximate $\mu(\tilde{Y}^{(i)})$ with

$$\hat{\mu}(\tilde{Y}^{(i)}) \triangleq \frac{1}{N} \sum_{j=0}^{N-1} K_G(\tilde{Y}^{(i)}|Y^{(j)}) \quad (14)$$

If we desire to approximate expectations under $\pi(x)$ of the form (2), we can simply use the *importance sampling* estimator

$$\hat{I} \triangleq \frac{\sum_{i=0}^{N-1} h(\tilde{Y}^{(i)}) \frac{\tilde{\pi}(\tilde{Y}^{(i)})}{\hat{\mu}(\tilde{Y}^{(i)})}}{W} \qquad (15)$$

where

$$W \triangleq \sum_{i=0}^{N-1} \frac{\tilde{\pi}(\tilde{Y}^{(i)})}{\hat{\mu}(\tilde{Y}^{(i)})} \qquad (16)$$

The advantage is that we also have an estimator for the partition function $Z(\beta)$: $\hat{Z}(\beta) \triangleq \frac{1}{N} W$ and not merely the *ratio* of two constants as often arises. The reason is that by construction, the normalization constant of $\mu$ is unity. The estimator for $\mu$ is $O(N)$ for each of the $N$ points; fortunately this overall $O(N^2)$ cost is quite reasonable since it is only done once, in the final stage of the algorithm. Note that one could also use a combination of Gibbs kernels at a *set* of temperatures as well.

## 6 Experiments

We present two sets of numerical results. The first set was primarily to establish that the method seems to work correctly. A 25 node fully-connected, binary valued, undirected graphical model under 6 values of the inverse temperature $\beta$ was used. The task was to estimate the $\log(Z(\beta))$ for each of these $\beta$. The system is an instance of an *Ising Spin Glass,* a model that assumes the form of the *Boltzmann machine* [19] in the AI community. Each component of the state $x_i \in \{-1, 1\}$. The energy function is given by:

$$E(x) = -\frac{1}{\sqrt{M}} \sum_{i<j} J_{i,j} x_i x_j \qquad (17)$$

where the $J_{i,j}$ are distributed according to the zero-mean, unit variance Gaussian.

The fact that the interactions $J_{i,j}$ are allowed to be of mixed sign makes this a very difficult and complex model; merely finding the global minimum is $NP$-Hard in general. For the small model here, $\log(Z(\beta))$ was calculated by brute force for each $\beta$. The estimates of $\log(\hat{Z}(\beta))$ from the LFIS methodology were compared against Sequential Monte Carlo [6] (SMC) with annealed distributions and Loopy Belief Propagation, and the results are presented in Table 1. 50 runs of LFIS and SMC were performed at each $\beta$; the "error" in Table 1 is defined as $|\log(\hat{Z}(\beta)) - \log(Z(\beta))|$. LFIS drew 1000 samples from sequences of 1000 flips each. For SMC 1000 samples were also used, and a linear cooling schedule was employed such that the number of computational steps matched those taken by the LFIS method (about 5000 annealing steps in our implementation.) SMC gives very good estimates of $\log(Z(\beta))$ of this system for low $\beta$ (high temperature,) clearly being the winning method for $\beta = 0.5$.

For $\beta = 1, 2$, LFIS and SMC perform comparably well (we must remember that this is a set of only 50 runs.) As $\beta$ rises further, LFIS begins to dominate. An interesting trend with SMC is the almost monotonically increasing error and variance with increasing $\beta$; in contrast, LFIS exhibits a small rise in error from $\beta = 0.5$ to $\beta = 2$, but it then declines sharply. Loopy BP gave respectable results for $\beta = 0.5$ and $\beta = 1$ but very large errors for the remaining $\beta$.

The previous example is too small to draw any concrete conclusions from, serving mostly as a demonstration of correctness of the LFIS against exact results, but the diverging accuracies of SMC and LFIS methods for decreasing temperatures certainly inspired us to investigate this effect further for large models. Unfortunately it becomes impossible to compare against the exact answers; nonetheless, if we force $\beta$ to be high, we know that under the equilibrium distribution, the system will primarily be in its low-energy states. As we mentioned in the introduction, the primary source of variance in Sequential Monte Carlo methods such as SMC is a failure of the proposal process to effectively sample the target distribution; the other source, due to importance sampling in a growing state space, can be effectively managed using strategies such as resampling [6]. Our compromise test methodology for comparing the LFIS against the annealing-based proposal process used in SMC is to generate a set of states using each method and to examine their properties under the target at $\beta$.

For this set of experiments we considered it very important to be thoroughly fair in terms of computational parity; we thus implemented an *event-driven* annealing (EDA) process of the type described previously. The process was such that at each step, one variable was flipped. The final state after the T EDA iterations was considered to be the sample generated by that run. The computational cost per iteration is almost identical to that of the LFQGS, except that latter needs to do slightly more bookkeeping in setting and unsetting the allowable moves. In our implementation, this translated to the LFQGS of $T$ steps running about 3% more slowly than the EDA process of the same length, although this can certainly be diminished in a less prototypical implementation. Note that the plain NFW at this temperature would perform very poorly.

Two large models were considered: the first, called 1000FC, was a 1000 variable fully-connected system whose energy had the form of Equation (17). Again, $J_{i,j}$ were drawn from the standard Gaussian. The second, called CUBE, was a binary-valued graphical model of three-dimensional lattice topology with dimension $4 \times 4 \times 16$. The interactions can be seen to be a special case of (17), where all $J_{i,j} = 0$ except those connecting neighbors on the lattice. The values of the remaining $J_{i,j}$ were 1 or $-1$ with equal probability instead of being Gaussian. Normalization and optimization on this topology is also $NP$-Hard [1]. $\beta$ was set to 20 for problem instances.

| $\beta$ | Exact $\log Z(\beta)$ | LFIS Error | Variance | SMC Error | Variance | LBP Error |
|---|---|---|---|---|---|---|
| .5 | 18.96 | $1.3 \times 10^{-3}$ | $1.0 \times 10^{-4}$ | $8 \times 10^{-4}$ | $2.2 \times 10^{-5}$ | 0.005 |
| 1 | 23.69 | $1.6 \times 10^{-3}$ | $7.9 \times 10^{-4}$ | $1.2 \times 10^{-3}$ | $1.6 \times 10^{-4}$ | 0.052 |
| 2 | 38.55 | $2 \times 10^{-3}$ | $3.7 \times 10^{-4}$ | $1 \times 10^{-3}$ | $6.7 \times 10^{-4}$ | 4.6 |
| 5 | 90.13 | $6 \times 10^{-4}$ | $1.3 \times 10^{-5}$ | $6.7 \times 10^{-3}$ | $3.2 \times 10^{-3}$ | 22.75 |
| 10 | 178.34 | $5 \times 10^{-4}$ | $9.0 \times 10^{-6}$ | $1.1 \times 10^{-2}$ | $3.5 \times 10^{-2}$ | 0.26 |
| 20 | 355.55 | $5 \times 10^{-4}$ | $3.2 \times 10^{-7}$ | $7.6 \times 10^{-2}$ | 0.2295 | 12.81 |

Table 1: The exact values $\log(Z(\beta))$ and of average error over 50 runs of the estimates $\log(\hat{Z}(\beta))$ for a 25 node complete graphical model using the 3 methods described in the text. Variances of $\log(\hat{Z}(\beta))$ are also included for LFIS and SMC; loopy BP does not have a variance as it is a deterministic method.

| Method | CUBE Mean energy | Energy Var | FC1000 Mean energy | Energy var |
|---|---|---|---|---|
| LFQGS | -401.52 | 0.899 | -751.04 | 2.02 |
| EDA | -393.54 | 25.281 | -737.07 | 46.19 |

Table 2: Results of the LFIS methodology against an EDA. The mean energies and variances are those obtained from 100 runs of each process. Note that at the system $\beta = 20$, the average state generated by LFIS is $2.65 \times 10^{121}$ more likely than one generated by EDA on FC1000 and $2.06 \times 10^{69}$ more likely on CUBE.

LFQGS and EDA were both run 100 times on each problem. 50000 and 100000 steps were taken by both on CUBE and 1000FC respectively. EDA employed a $\beta$ schedule which ascended linearly from $\beta = 0.001$ to $\beta = 20$ in the alloted time steps. For LFQGS, the LF move size range was set to $\Gamma_{min} = \lfloor M/8 \rfloor$ and $\Gamma_{max} = \lfloor M/6 \rfloor$. Table 2 displays the results of these simulations; the average energy of a sampled state over the runs is shown for the two methods on both models; the variance in the energies is also shown. It is clear that on these large models, for the given computational resources, the samples generated by our methodology are of far higher quality than those obtained even by using the sophisticated implementation of annealing. The average sample generated by our method from FC1000was $2.65 \times 10^{121}$ times more likely at $\beta$ than one generated by EDA; for the CUBE model, that ratio is $2.06 \times 10^{69}$. The samples generated with EDA also show a very large spread of energy. The SMC procedure that uses these samples is doomed to failure; the reweighting step in it cannot save a particular population that has not visited the right regions, it can only save one that has visited the right ones with the wrong frequencies.

# References

[1] F Barahona. On the computational complexity of Ising spin glass models J. Phys. A: Math. Gen. 15 3241-3253, 1982.

[2] D. Blei, A. Y. Ng, and M. I. Jordan. Latent Dirichlet allocation. *J. Mach. Learn. Res.*, 3:993–1022, 2003.

[3] A.B. Bortz, M.H.Kalos, J.L. Lebowitz A new algorithm for Monte Carlo simulation of Ising spin systems. *Jour. Computat. Phys.*, 17:10–18, 1975.

[4] P. Carbonetto and N. de Freitas. Conditional mean field. *NIPS*, 2006.

[5] J. Cheng and M. J. Druzdzel AIS-BN: An Adaptive Importance Sampling Algorithm for Evidential Reasoning in Large Bayesian Networks. *Jour. Art. Intell. Res.*, 13, 155-188, 2000.

[6] P. del Moral, A. Doucet, and A. Jasra. Sequential Monte Carlo samplers. *J. Roy. Statist. Soc., Ser. B*, 68:411–436, 2006.

[7] L. Devroye *Non-Uniform Random Variate Generation.* Springer-Verlag, 1986.

[8] R. Fung, K. Chang. Weighting and integrating evidence for stochastic simulation in Bayesian networks. UAI, 1989. Morgan Kaufmann.

[9] G. Getz, N. Shental and E. Domany. Semi-Supervised Learning – A Statistical Physics Approach. In *ICML Workshop on Learning with Partially Classified Training Data*, 2005.

[10] V Gore and M Jerrum. The swendsen-wang process does not always mix rapidly. In *29th Annual ACM Symposium on Theory of Computing*, 1996.

[11] F. Hamze and N. de Freitas Large-Flip Importance Sampling Univ. of. Brit. Columb. Comp. Sci. Tech. Rep, 2007.

[12] Y. Iba. Extended ensemble Monte Carlo. *Int. Jour. Mod. Phys.*, C12:623–656, 2001.

[13] C Jarzynski. Nonequilibrium equality for free energy differences. *Phys. Rev. Lett.*, 78, 1997.

[14] S Kirkpatrick, C D Gelatt, and M P Vecchi. Optimization by simulated annealing. *Science*, 220:671–680, 1983.

[15] J D Lafferty, A McCallum, and F C N Pereira. Conditional random fields: Probabilistic models for segmenting and labeling sequence data. In *ICML*, 2001.

[16] J. Liu, J. L. Zhang, M. J. Palumbo and C. E. Lawrence. Bayesian clustering with variable and transformation selection. In *Bayesian Statistics*, 7:249–275, 2003.

[17] M E J Newman and G T Barkema. *Monte Carlo Methods in Statistical Physics*. Oxford University Press, 1999.

[18] J. D. Munoz, M. A. Novotny and S. J. Mitchell. Rejection-free Monte Carlo algorithms for models with continuous degrees of freedom. *Physical Review Letters*, 67(2), 2003.

[19] D. Ackley, G. Hinton, T. Sejnowski. A learning algorithm for Boltzmann machines. *Cognitive Science*, 9, 147-169, 1985.

[20] F. Glover. Tabu Search - Part I. *ORSA Journal on Computing*, 1(3), 190-206, 1989.

[21] K. Smyth, H. H. Hoos, T. Stützle Iterated Robust Tabu Search for MAX-SAT Canadian Conference on AI 2003: 129-144.

[22] E.D. Taillard. Robust taboo search for the quadratic assignment problem. Parallel Computing, 17:44, 1991.

[23] R H Swendsen and J S Wang. Nonuniversal critical dynamics in Monte Carlo simulations. *Phys. Rev. Let.*, 58(2):86–88, 1987.

[24] S. S. Tham, A. Doucet, and K. Ramamohanarao. Sparse Bayesian learning for regression and classification using Markov chain Monte Carlo. In *ICML*, pages 634–641, 2002.